\documentclass[11pt]{article}
\usepackage{epsfig}
\def\lvm{\leavevmode\hbox to\parindent{\hfill}}
 
\def\fun#1#2{\lower3.6pt\vbox{\baselineskip0pt\lineskip.9pt
\ialign{$\mathsurround=0pt#1\hfil##\hfil$\crcr#2\crcr\sim\crcr}}}
\def\plotone#1{\epsfysize=15cm\epsffile{#1}}

\title{Gamma-ray Bursts, Type Ib/c Supernovae  and \\
       Star-forming Sites in Host Galaxies}
\author{D.Yu.Tsvetkov$^{1}$, S.I.Blinnikov$^{2,1}$, N.N.Pavlyuk$^{1}$}
\date{}

\begin{document}
\begin{titlepage}
\maketitle
\thispagestyle{empty}
\vskip 1cm
\centerline{${}^1$\it Sternberg Astronomical Institute, 119899 Moscow,
  Russia}
\centerline{e--mail: tsvetkov@sai.msu.su;
blinn@sai.msu.su; pavlyuk@sai.msu.su}
\centerline{${}^2$\it Institute for Theoretical and Experimental Physics,
  117218 Moscow, Russia}
\centerline{e--mail: sergei.blinnikov@itep.ru}

\begin{abstract}
  The data on the location of gamma-ray bursts (GRBs)
relative to their host galaxies are used to derive the distribution
of surface density of GRBs along the galaxy radius. It is shown
that the gradient of GRB surface density changes abruptly near
the half-light radius.
In the central parts of galaxies the
distribution of GRBs resembles closely the luminosity distribution,
while in the outer parts the galactic surface brightness falls
much steeper than the GRBs density.
The radial distribution of type Ib/c
supernovae is investigated on the basis of enlarged statistics.
It is shown that SNe Ib/c do not differ significantly from other types
of supernovae and their distribution is more similar to the one
for recent star formation sites than that of GRBs.
In spite of
the poor statistics of GRBs, the difference in the distributions of
active star formation regions and GRBs appears to be significant.
We get the Kolmogorov-Smirnov probability $P_{\rm KS}$ of only 4\%
that GRBs and star-forming sites belong to the same distribution.
The correlation of GRBs with the distribution of dark matter in the
outer parts of galaxies is not excluded.
\end{abstract}

{\sl Keywords:}
Gamma-rays: bursts --- Supernovae --- Dark Matter

\end{titlepage}

\vspace{0.7cm}

\section*{Introduction}\lvm
\label{intr}
\addtocounter{page}{1}

The investigations  of  spatial distribution of supernovae (SNe)
in parent galaxies have given important clues to the probable
nature of progenitors, whose evolution results in
SN outbursts of various types (see, for example, Bartunov et al.,
1992; Bartunov et al. 1994; Bartunov, Tsvetkov 1997; van
den Bergh 1997).

The observational data on gamma-ray bursts (GRBs) available now are
somewhat similar to the data for SNe.
Certainly, it is necessary to remember that the errors of GRB localization
in the host galaxies are larger than for supernovae,
and statistics of localized GRBs is much poorer.
Nevertheless, the technique used in the papers by Bartunov et al.(1992, 1994)
has given important results for SNe, when their statistics
was also not very rich.
Therefore, it is of interest to apply the technique used for SN studies
to the research of the GRB distribution.

Discovery of GRB afterglows and the measurement of their redshifts
puts stringent requirements on theoretical models of gamma-ray bursts
which are extremely hard to satisfy
(see reviews by  Piran 1999; Postnov 1999).
The  GRB distribution in parent galaxies is very important for
comprehension of their nature.
Models, in which the gamma-ray burst is born at a blow-up of massive stars
should result in a correlation of GRBs with star-forming regions.
In the model of merged double relativistic stars the gamma-ray burst
can take place far away from the sites of origin of massive stars.
And at last, the GRB models involving exotic versions of dark matter (see
the review by Blinnikov, 2000) are offered.
Finding a correlation of the GRB distribution with any type
of the stellar population or with dark matter will allow one
to shed light on their origin.

\vspace{0.7cm}
\section*{Investigating the GRB distribution} \lvm

The recent observational data on the location of GRBs in parent galaxies
are assembled in the  paper by Bloom et al. (2000).
For 20 GRBs the angular
distances from centers of galaxies are determined, among
them for 15 GRBs the redshift is known, and linear distances
are determined as well.
For normalization
of distances, the effective radius of a galaxy $R_{\rm eff}$
is used, i.e. the  radius inside which half of the luminosity is radiated.
The effective radii are determined directly from
observations or are estimated on the basis of an empirical
relation between the galaxy stellar magnitude and the effective radius.

Bloom et al.(2000) have constructed cumulative distributions for relative
(angular) and absolute
(physical)  GRB offsets from centers of their apparent host galaxies,
and also a histogram of the offset distribution as a function of
relative distances with account of errors of measurements.
Bloom et al.(2000) conclude, that the observed distributions are in
good agreement with the distribution of the population belonging
to the exponential disk of galaxies, and does not agree with
the distribution of  merging relativistic objects, i.e. binary neutron stars
and neutron stars in pairs with black holes.
Those pairs can travel a long distance from the sites of massive star
formation before they collapse and produce a GRB.

In our opinion, the representation of radial distribution of objects as
a function of surface density from distance to center is preferable
for comparing with the distributions of various populations in galaxies.

As in the paper by Bartunov et al.(1992), we have determined the
smoothed surface density of GRBs,
$\sigma_i = N_i/\pi (r_{i + 1}^2 - r_i^2)$, where $r_i$ is the
radius of the $i$-th bin expressed either in terms of effective
radius of a galaxy ($r_{\rm rel}$) or in kiloparsecs;
$N_i$ is the smoothed number of objects in the $i$-th bin computed
from true numbers $n_i$ as $N_i=0.5n_i +0.25(n_{i-1}+n_{i+1})$.

In  Fig.~1 the dependence of the surface GRB density
on the relative distance to center (in terms of effective radius) is shown.
The data for 18 GRB are used: two of the 20 GRBs with relative distances 9.7
and 11.0 are excluded from the list. For them, according to
Bloom et al.(2000) the probable host galaxies are not identified,
and the indicated values of offsets are assigned to the nearest
galaxies in the field.
The interval of the binning is taken equal to 0.4;
only the statistical errors, proportional to $(N_i)^{0.5}$ are shown.

It is clear, that the gradient of the surface density of GRBs,
$d\lg(\sigma) / d r_{\rm rel} $ changes sharply at $r_{\rm rel} = 1.0$
-- the steep decreasing of density is replaced by a slower decline.
The dependence of $\lg(\sigma)$ on $r_{\rm rel}$ can be approximated
by two straight lines, with the gradients
$ -1.4 \pm 0.1 $ and $ -0.28 \pm 0.02 $ respectively.

In the papers by Bartunov et al.(1992), Bartunov and Tsvetkov (1997) the
gradients of logarithms of surface density for supernovae of various types
and, for comparison, also for various types of the stellar population
in galaxies were determined.
For normalization, the values of diameters of galaxies $D_{25}$
from the catalogue RC3 (de Vaucouleurs et al. 1991) were used.
To compare
these data to the results for GRBs, we have calculated the mean
ratio $D_{25}/2R_{\rm eff} = 2.7$ for all galaxies in RC3,
for which these parameters are known.
Then, for the normalization of distances on the radius $0.5D_{25}$, the
gradients of $\lg(\sigma)$ will make $ -3.8 $ and $ -0.76 $ respectively.
The density gradient of GRBs in central areas of galaxies is essentially
steeper than for the studied supernova samples, and it agrees well
with the gradient of surface brightness of elliptic galaxies.
However, in external areas of galaxies
the gradient of the surface GRB density  is more shallow than
for supernova and for surface brightness of galaxies of all types,
though the number of GRBs in this area is not sufficient
for firm conclusions.
The distribution of star-forming regions
in spiral galaxies differs strongly from the GRB distribution
both in central areas, and in periphery of galaxies.
In Fig.1 the distributions of surface brightness in elliptical and
spiral galaxies are shown for comparison, and also
the distribution of OB-associations in M33 and H~II regions
in NGC 3184 according Burstein (1979), Humphreys, Sandage (1980),
Boroson (1981), Hodge, Kennicutt (1983).
These distributions were
shifted along the ordinate for the best adjustment with the
data for GRBs.
It is clear that the GRB distribution in central areas of galaxies
agrees  well with the distribution of the luminosity both in elliptical
and in spiral galaxies.
The distribution of star-forming regions differs sharply from the
GRB surface density.

Fig.~2 demonstrates the dependence of the GRB surface density
on the distance to centers of galaxies expressed in kiloparsecs.
The data for 14 GRB and the binning interval of 1.2 kpc are used.
For $ r < 5 $ kpc the dependence is well approximated by a straight line,
the gradient $ \lg (\sigma) $ is equal to $ -0.40 \pm 0.03 $,
this corresponds to the value
$r_0=1.1$ kpc in the expression $ \sigma \sim {\exp} (-r/r_0) $.
As well as for the distribution on the relative offset, the GRB density gradient
here is significantly steeper than for all studied supernova samples.
The dependence of the GRB surface density on radius is in very good
agreement with the distribution
of surface brightness in elliptical galaxies and differs from that
of star-forming regions.

The GRB distributions considered above differ from radial distributions of
supernovae of all types (Bartunov et al. 1992; Bartunov, Tsvetkov 1997;
van den Bergh 1997).
The supernova distribution is represented rather well by an exponential law
of the density, decreasing with a constant factor, practically in the whole
disk of spiral galaxies, which matches well the typical distribution
of luminosity in the disk of spiral galaxies.
A special interest represents the research of type Ib/c supernova
distribution: first, Bartunov et al.(1992)  found an unexpectedly
strong concentration to centers of galaxies for them; second,
some of GRBs are claimed to be physically related to SN Ib/c outbursts.

For our study, the data on 50 SNe Ib/c from the SAI supernova Catalogue
(Tsvetkov et al., 2000) were used.
The obtained radial distributions of SNe Ib/c are also shown in
Figs.1 and 2.
It is clear, that the dependence of a log of surface density
SN Ib/c both on the relative offset and on the physical radial distance
is represented rather well by straight lines.
The gradient of $\lg(\sigma)$ is $ -2.1 \pm 0.1 $
when the offset is expressed in units of $0.5D_{25}$, and the
dependence of $\lg(\sigma)$ on the physical radius in kiloparsecs
makes the gradient $ -0.18 \pm 0.01 $ with $ r_0 = 2.4 $ kpc.
The distributions of SN Ib/c are in a reasonably good
agreement with the distributions of OB-associations and H~II regions and
differ noticeably from the GRB distributions.
The density gradient of SN Ib/c practically coincides with the
gradient of surface brightness of spiral galaxies in their external parts
and is close enough to the appropriate data for supernovae of other types.

The qualitative conclusions drawn above are confirmed by the results of
applying the Kol\-mo\-go\-rov-Smirnov test  (Press et al. 1986)
to the studied samples of objects.
The GRB distribution over the relative offset was compared to
the other  distributions shown in Fig.1.
We get the following values of the probability $P_{\rm KS}$
that a given pair belongs to the same distribution.
For the pair
\begin{itemize}
\item [](GRB - surface brightness of spirals): $P_{\rm KS} =$68\%;
\item [](GRB - surface brightness of ellipticals): $P_{\rm KS}=$40\%;
\item [](GRB - OB-associations and H~II regions): $P_{\rm KS}$=4\%;
\item [](GRB - supernovae of type Ib/c): $P_{\rm KS}$= 9\%.
\end{itemize}

\vspace{0.7cm}
\section*{Discussion} \lvm

The obtained GRB offset distributions relative the host galaxies
already allow us to draw some conclusions.
For a while they remain preliminary, but with
the accumulation of the observational data and the improvement
of statistics, the technique used in this paper will be very
helpful for revealing the nature of gamma-ray bursts.

If a gamma-ray burst is born during collapse of a massive star
(Woosley 1993;  Fryer, Woosley 1998; Fryer et al., 1999a, 1999b;
Popham et al. 1999; Gershtein 2000; Cherepashchuk, Postnov  2000),
a strong correlation of GRBs with star-forming areas
or with Wolf-Rayet stars should be observed.

In the model of merging binary neutron stars (Blinnikov et al. 1984;
Paczy\'nski 1986; Eichler et al. 1989) and
pairs of neutron stars with black holes (Lattimer, Schramm 1974,1976),
a gamma-ray burst  can occur on a considerable distance
from the areas of origin of massive stars, although this is only one option
in the merger scenario, and the problem demands a special research.

And at last, because of numerous difficulties of explaining GRBs in
the standard physics, the models are offered that involve exotic particles,
various versions of dark matter etc.
(Loeb 1993; Bertolami 1999; Demir, Mosquera Cuesta 1999;
Iwazaki 1999; Blinnikov 1999; Berezhiani, Drago 2000;
see the review by Blinnikov 2000).

The nature of dark matter remains obscure.
We touch only briefly upon some of the recent results in this area
which can be relevant for the GRB problem.
Until recently, hypothetical very weakly interacting particles
were considered as the most probable candidates for the explanation of
all known data on dark matter (for example, superlight axions or,
on the contrary, very massive
WIMPs - Weakly Interacting Massive Particles, such as neutralinos).
These particles are non-relativistic in the present epoch
(i.e. they are probable constituents of ``Cold Dark Matter" = CDM),
this is necessary for explaining the formation of galaxies
and their clusters.
However,  the theory using such particles, has met with a number
of difficulties.
At first, the rotation curves of galaxies, which are also determined by
dark matter, show that CDM should have constant density
in central areas of galaxies, whereas the calculations in CDM models
for weakly interacting particles produce structures with a sharp cusp
in the center and a density decrease according the law $\propto 1/r^2$.
Secondly, the calculations give too large number of CDM clouds in massive
halos of galaxies -- a factor of 10 - 50 more than permissible.

Spergel and Steinhardt (2000) suggested that both these
problems can be solved if the dark matter particles
sufficiency strongly and elastically interact among
themselves: central galactic peaks are flattened, and the
inter-galactic clouds dissipate.

In the paper by Meneghetti et al. (2000) the limits on the strength
of the self-interaction of CDM particles are obtained.
It is found, that when the cross-section of interaction of the
dark matter particles is too large, the structure of clusters of galaxies
is flattened so strongly, that it is impossible to explain such
observable phenomena as giant arcs in the images of distant quasars
and galaxies created by invisible matter of clusters due to the
effect of the gravitational lensing.
The comparison with observations shows that the cross-section
of interaction of CDM particles per unit mass should not exceed
$\sim 0.1$ cm$^2$ g$^{-1}$.

It is interesting, that the upper limit on self-interaction
follows also from other reasons.
Burkert (2000) has presented the review of results indicating
that, when the cross-section is too large, the homogeneous isothermal
CDM core collapses quickly.
It is important, that the cross-section can not be much less.
Otherwise, one fails to achieve the desirable effect of smoothing
the galactic cusps.

The large cross-section of interaction can result in the formation of
stellar-like objects consisting of dark matter, which can collapse and
generate gamma-ray bursts via various paths (Blinnikov 2000).
One should not forget, that the dark matter can consist of
diverse particles, e.g. in the models of `mirror' matter
(Kobzarev et al. 1966), and an hierarchy of the stellar populations
can be formed there, similar to ordinary matter (Blinnikov, Khlopov 1983;
Berezhiani et al. 1996, Foot, Volkas 1997,2000; Silagadze 1999;
Mohapatra, Teplitz 2000).
By the way, this can explain MACHO events (Berezhiani et al. 1996;
Blinnikov 1998; Silagadze 1997; Foot 1999; Mohapatra, Teplitz 1999).

The invisible matter in many spiral galaxies can be distributed as
neutral hydrogen in central areas, and in the outer parts its density
falls considerably slower (Hoekstra et al. 2000).
The analysis of observations, carried out by us, shows that the
radial GRB distribution  in internal
parts of galaxies is similar to the population of a spherical
population of galaxies, and in external parts their density drops slower,
than for visible disks of spiral galaxies, i.e. it is qualitatively
similar to dark matter.

Our results show, that GRBs are distributed along the radii unlike
the sites of active  star formation in nearby galaxies.
However, many observers of GRB afterglows insist that they see them
mainly in galaxies totally covered by a burst of
star formation (Sokolov et al. 2000).
The morphology of these galaxies is very irregular.
In such a case it would be inappropriate to compare these peculiar
galaxies with nearby normal spirals.
Nevertheless, in these cases again it is impossible to say that
a gamma-ray burst is necessarily produced by an ordinary massive star.
Something should cause bursts of star formation in such galaxies.
Sometimes, but it is very rare, star-forming galaxies have satellites,
whose tidal force could serve as the ``trigger" for bursts of
stellar births.
However, mostly they seem to be single (see Taylor et al. 1995;
Telles, Maddox 2000).
It is very probable, as noted by Trentham et al. (2000)
(and even earlier by Blinnikov 1999, 2000), that such galaxies
interact gravitationally with a satellite consisting of dark matter.
If GRBs are produced by dark matter objects, we should
see a natural correlation with a star formation burst in visible matter,
but certainly, it would not be a proof of genetic connection of gamma-ray
bursts and ordinary massive stars.

Of course, one should not forget also that there are bright afterglows
like GRB~000301C, and no galaxies with brightness down to $29^m$
are revealed there.
There are also other hints on existence of absolutely dark galaxies
(Trentham et al. 2000).

\vspace{0.7cm}

{\bf Conclusions.}

GRBs are distributed like visible matter in bulges of spiral and
elliptic galaxies, and in the outer parts
their surface density falls much more slowly than the luminosity
of galaxies.
They do not follow the distribution of
the regions of active  star formation or SNe Ib/c in nearby spiral
galaxies, but follow their spherical population in central parts.
The current data do not at all exclude the GRB correlation with  dark
matter in the outskirts of galaxies
(in spite of a powerful selection effect:
the afterglows should be more often observed in the regions with
the higher density of ordinary matter).

\vspace{0.7cm}

{\bf Acknowledgements.} We thank S.S.Gershtein, K.A.Postnov,
M.E.Prokhorov and V.V.So\-ko\-lov for stimulating discussions and
O.K.Silchenko for invaluable help.
The work is supported by the grant RFBR 99-02-16205.



\let\jnlstyle=\rm
\def\refjnl#1{{\jnlstyle#1}}
\def\aj{\refjnl{AJ}}
\def\araa{\refjnl{Ann. Rev.  Astron. Astrophys.}}
\def\apj{\refjnl{Astrophys. J.}}
\def\apjl{\refjnl{Astrophys. J. Lett.}}
\def\apjs{\refjnl{Astrophys. J. Suppl.}}
\def\ao{\refjnl{Appl.~Opt.}}
\def\apss{\refjnl{Astrophys.Space Sci.}}
\def\aspph{\refjnl{Astroparticle Phys.}}
\def\aap{\refjnl{Astron.Astrophys.}}
\def\nat{\refjnl{Nature}}
\def\prd{\refjnl{Phys.~Rev.~D}}
\def\yaf{\refjnl{Sov.J.Nucl.Phys.}}

\bigskip

\begin{figure}
\centering{
\plotone{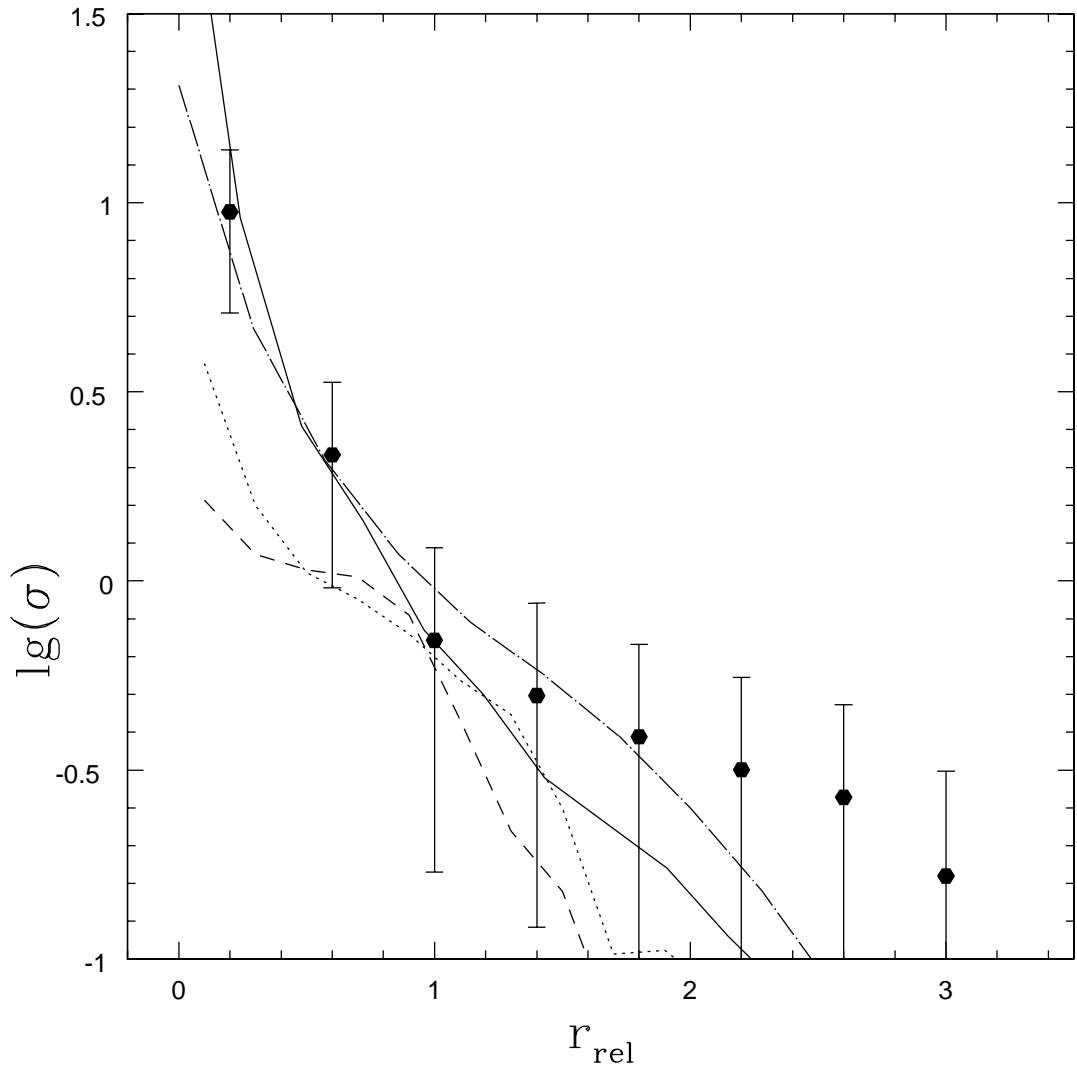}
            }
\caption{Dependence of logarithm of the GRB surface density
on relative offset (in units of effective galaxy radius).
Solid dots correspond to the surface density of GRBs with respect to
the center of each binning interval, solid line is the dependence of surface
brightness of ellipticals on radius, dashed-dotted --
the same dependence for spirals,
dashed -- OB-associations in M33 and
H~II zones in NGC 3184, dotted line -- the distribution of type Ib/c
supernovae.}
\label{fig1}
\end{figure}

\begin{figure}
\centering{
\plotone{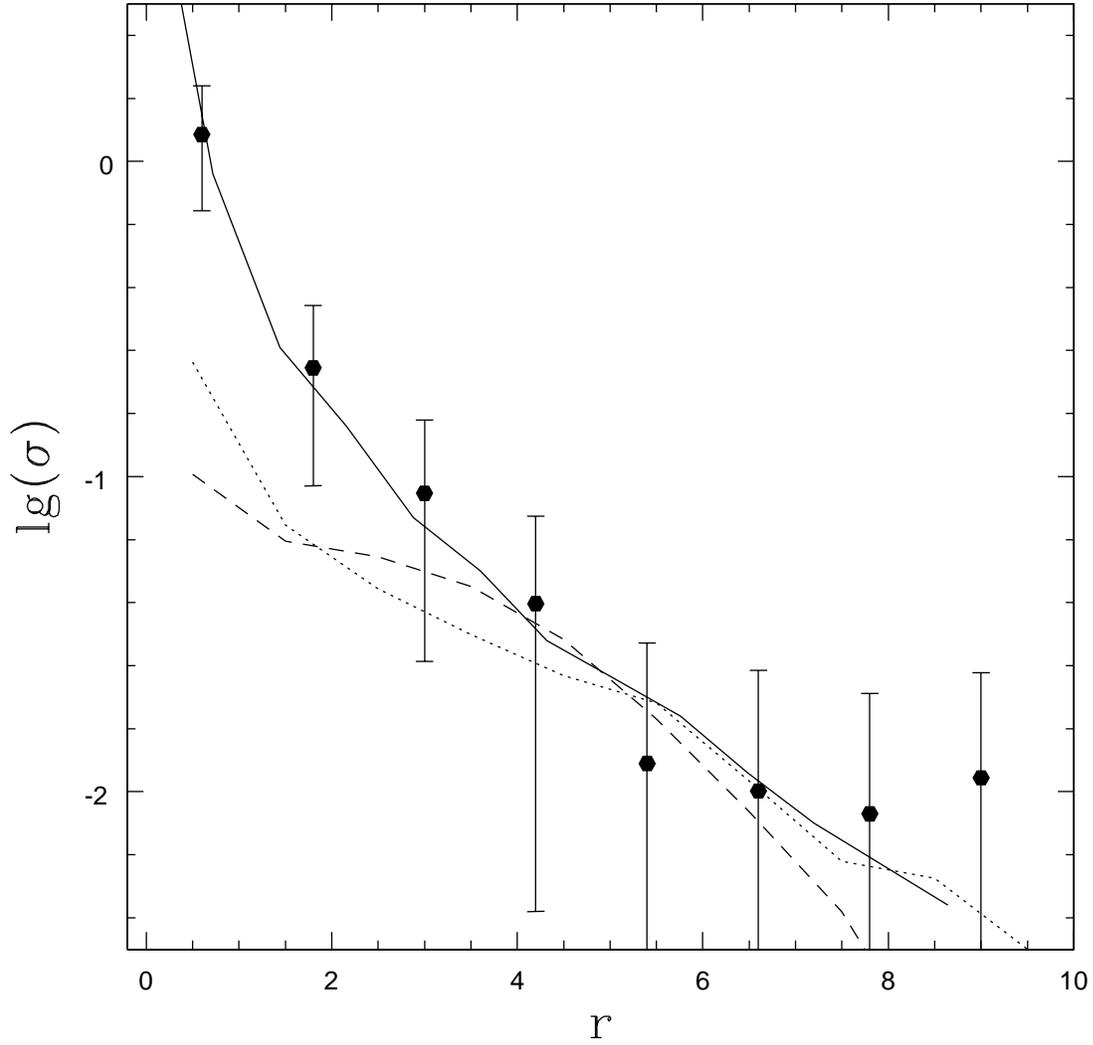}
            }
\caption{Dependence of the surface density of GRBs (solid dots) on absolute distance,
in units of kpc.
Solid line corresponds to the dependence
of surface brightness of ellipticals on radius,
dashed -- OB-associations in M33 and H~II zones in NGC 3184,
dotted line -- the distribution of type Ib/c supernovae.}
\label{fig2}
\end{figure}

\end{document}